# Increasing the sensitivity of a Shack-Hartmann sensor

MARKUS KASPER[1], DOUGLAS LOOZE[2], STEFAN HIPPLER[1], RIC DAVIES[3], ANDREAS GLINDEMANN[4]

[1]Max-Planck-Institut für Astronomie , [2]University of Massachusetts,
[3]Max-Planck-Institut für extraterrestrische Physik, [4]European Southern Observatory

**Abstract**. The Max-Planck institutes for astronomy (MPIA) and for extraterrestrial physics (MPE) run an adaptive optics (AO) system with a laser guide star at the 3.5 m telescope on Calar Alto, Spain. This system, called ALFA, produces now scientific results and works excellent with natural guide stars (NGS) as faint as 13th magnitude in R-band. The ultimate goal however is to achieve similar performances with the laser guide star (LGS) which is faint and extended. We introduce the Shack-Hartmann wavefront sensor implemented in ALFA and present our efforts in increasing its sensitivity by using advanced centroiding and wavefront reconstruction algorithms.

## 1. Introduction

The correction of wavefront distortions caused by atmospheric turbulence require fast wavefront measurements of some spatial resolution. Both, the demands on speed and resolution, are set by the atmospheric conditions and require a certain guide star brightness in order to achieve an accurate estimate of the wavefront. For faint guide stars, this estimate will be based on noisy measurements.

The Shack-Hartmann sensor (SHS) is the commonly used device to measure wavefronts in adaptive optics. Its noise characteristics have been developed analytically for the use of the weighted pixel average algorithm to determine the spot centroids (Rousset, 1994). We simulated the accuracy of various centroiding algorithms under changing environmental parameters like detector read-noise and pixel-scale.

After having obtained the spot locations (wavefront slopes) various estimators can be used to reconstruct the incident wavefront. We compared the least squares and the weighted least squares estimators. The latter uses the knowledge of the noise covariance to provide a better estimation in the low signal case.

## 2. The ALFA Shack-Hartmann sensor

A Shack-Hartmann sensor subdivides the pupil into subapertures, and images each one separately. The differences between the locations of these subimages and the reference positions (obtained during calibration) represent local wavefront slopes of which the wavefront has to be reconstructed.

The ALFA-SHS was provided by Adaptive Optics Associates Inc., Cambridge, USA. The close-up view in Figure 1 displays the individual components (for a description of the whole system, see e.g. Hippler et al., Glindemann et al., or Kasper et al.).

The field select mirror is in a re-imaged telescope pupil plane, so that tilting this mirror results in a movement of the image on the SHS-detector. This mirror allows us to place a star within 30″ radius of the optical axis onto the SHS.

Two field stops and a reference fiber source can be inserted into the re-imaged focus. The field stops are required if the guide star is in a dense star cluster leading to cross talk between subimages, or if the laser guide star is used and the Rayleigh-scattered light has to be blocked.

Eventually, the telescope pupil is imaged onto a microlens array that produces the spot pattern. The lenslets can be changed remotely from a single lens to an array covering the pupil with 30 subapertures.

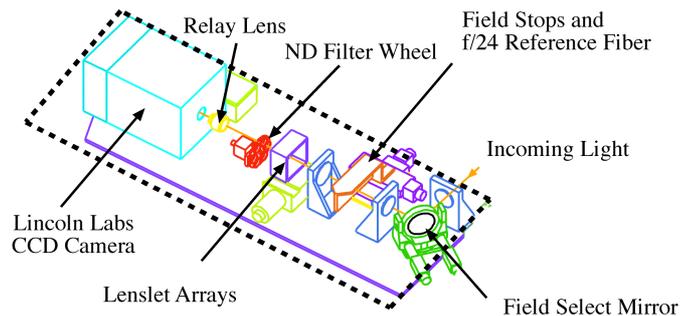

**Figure 1**: The ALFA Shack-Hartmann sensor

The spot pattern can be focused on the CCD camera via a relay lens. Since this camera is mounted on a motorised stage, we are able to adjust the pixel-scale (usually 0.5″ per pixel) and to accommodate the difference in focus between a natural guide star (NGS) at infinity and the laser guide star (LGS) at around 90 km.

## 2.1. The microlens arrays

There are 4 different microlens arrays mounted on a motorised stage, in order to react to various seeing conditions and guide star magnitudes. The arrays are coated on both sides to achieve maximum transmission.

Figure 2 shows the available arrays and their orientation to the telescope pupil.

The D3 array (D3 stands for 3 subapertures per pupil diameter) provides 6 subapertures over the telescope pupil, since the central lens is obscured by the secondary telescope mirror. This array is used for very faint guide stars.

There are two D5 arrays with different focal lengths providing 18 subapertures each. These arrays are used for faint guide stars and the LGS.

The D7 array provides 30 subapertures and is used for bright NGS.

Additionally, it is possible to use a single lens for LGS diagnostics like brightness, spot size etc.

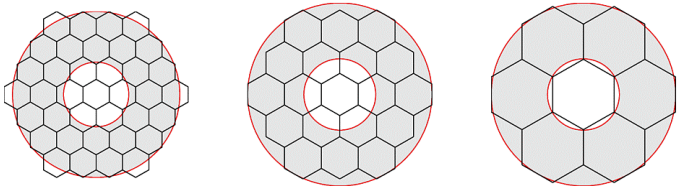

**Figure 2**. The ALFA microlens arays. *Left*: D7 array with 30 subapertures. *Middle*: D5 array with 18 subapertures. *Right*: D3 array with 6 subapertures.

## 2.2. The detector

The SHS detector is a thinned 64 x 64 pixel CCD that is used in frame transfer mode. The maximum frame rate is 1206 Hz with a pixel clock of 1.8 MHz and a read-noise of about 4 electrons. In figure 3, the quantum efficiency is shown to be around 80% between 600 nm and 750 nm. The underlying grey curve shows the quantum efficiency of the tip-tilt-tracker CCD (EEV39).

The camera has a two stage thermo-electric cooler providing an operating temperature of −35˚C which reduces the dark currents of the CCD. Thermal energy is transported outside the camera via a heat-pipe.

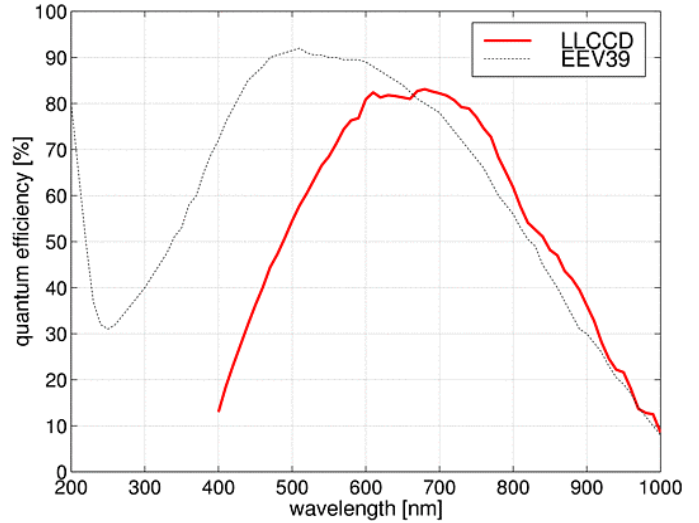

**Figure 3**. Quantum efficiency of the SH-CCD (LLCCD).

## 3. GS brightness and AO performance

Figure 4 was taken from Louarn et al. (1998) and shows the simulated dependence of the performance of an AO system (the Strehl-ratio is used as an indicator for image quality) from the guide star magnitude. Two main regimes are prominent:

❍ A fairly constant region for brighter guide stars. The performance is limited by fitting and aliasing errors which can't be avoided since they are inherent in the wavefront sensor and the chosen control modes.
❍ A quick drop in performance going to fainter guide stars. Inaccurate wavefront reconstruction because of measurement noise starts to dominate.

In principle, two possibilities exist to increase the flux in the subapertures:

❍ Bigger (but fewer) subapertures. This would result in a coarser sampling of the wavefront and larger fitting error because of less and erroneous reconstructed modes.
❍ Longer integration (lower framerate). This would produce a larger error because of temporal anisoplanatism.

The three X's in Figure 4 show the results obtained with ALFA in K-band and confirm the general behaviour.

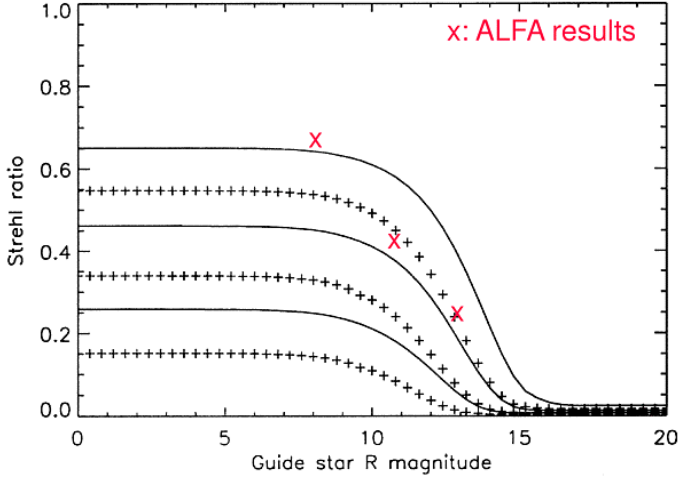

**Figure 4**. Simulated Strehl-ratio as a function of guide star magnitude (taken from Le Louarn et al. (1998)). *Lines*: Good seeing model. +: Median seeing model. *Up-Down*: K-, H-, and J-band. *X's*: ALFA results.

**PSF examples**

The left image in Figure 5 shows a 10s K-band exposure of an $m_V = 8.1$ star. The PSF has a Strehl-ratio of 72% and a FWHM of $\leq 0.14''$.

The image to the right is a 20s K-band exposure of an $m_R = 12.7$ star. The Strehl-ratio is around 20% and the FWHM is 0.16''. Some of the 2s exposures which have been added to obtain the long exposure have Strehl-ratios of more than 30%.

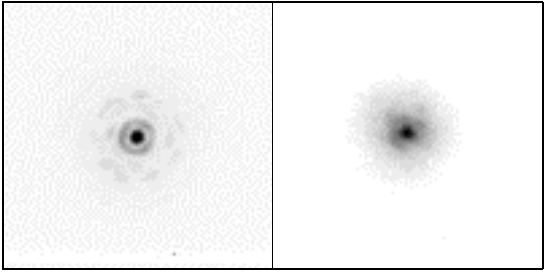

**Figure 5**: K-band long exposures corrected with ALFA. *Left*: NGS $m_V = 8.1$. *Right*: NGS $m_R = 12.7$

## 4. Centroiding algorithms

In order to improve the low signal performance we investigated different centroiding algorithms which have to work fast enough to be used in closed loop. Four algorithms have been studied:
- WPA. The weighted pixel average is calculated by $x_{wpa} = \sum_i x_i \cdot I_i / \sum_i I_i$.
- CWPA. The constrained WPA uses only 5x5 pixels around the brightest one.
- THRESH. The threshold WPA masks out all pixels with intensities below a certain value.
- WWPA. The weighted WPA raises all pixel values to the 1.5$^{th}$ power before doing the WPA. The motivation for this was to weight each pixel with its photon noise. ($I_{i,wwpa} = I_i \cdot \sqrt{I_i} = I_i^{1.5}$)

### 4.1. Simulations

We used the following Monte Carlo simulation process:
- Create arrays with read-noise (e.g. 10x10 pixels).
- Choose a large number of centroid-positions by random (e.g. 500 centroids for the upcoming simulations).
- Create spots by projecting gaussian functions onto the pixel grid.
- Add photon-noise to the pixel intensities.
- Convert electrons into counts.
- Apply algorithms to calculate centroids.
- Compare them with the random original positions.

**Centroiding vs. flux**

Figure 6 plots the simulated rms centroiding error in pixels against the flux level. Effects like saturating the camera are not taken into account in this case. The simulation parameters for these curves match the ALFA case (4 electrons read-noise, pixel-scale: 2 pixels/FWHM, array-size: 10x10 pixels)

The WPA is the least accurate algorithm, while the others deliver comparable performances. The WWPA has a slight advantage over the other algorithms around a signal of 1000 e$^-$, which resembles the faint guide star case.

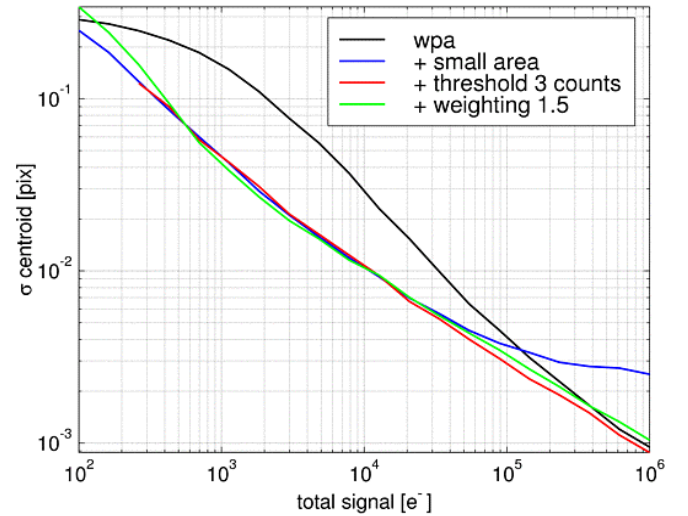

**Figure 6**. Comparison of the different centroiding algorithms.

The error variance is approximately proportional inversely proportional to the signal which is the analytical result.

**Centroiding vs. pixel-scale**

Figure 7 plots the simulated rms error for the spot centroids against the signal level for different pixel-scales. Other simulation parameters have been: 4 e$^-$ read-noise, WWPA algorithm, 10x10 pixel array-size. Note that the y-axis is given in some arbitrary length units because the pixel-size is variable.

A very coarse pixel-scale of 1 pixel / FWHM is only favourable for very low signals but saturates quickly for higher fluxes. Using a fine sampling of 3 pixels / FWHM has no advantage compared to 2 pixels / FWHM in the interesting region below a total signal of 10000 e$^-$ (it is in fact disadvantageous, i.e. oversampling should be avoided).

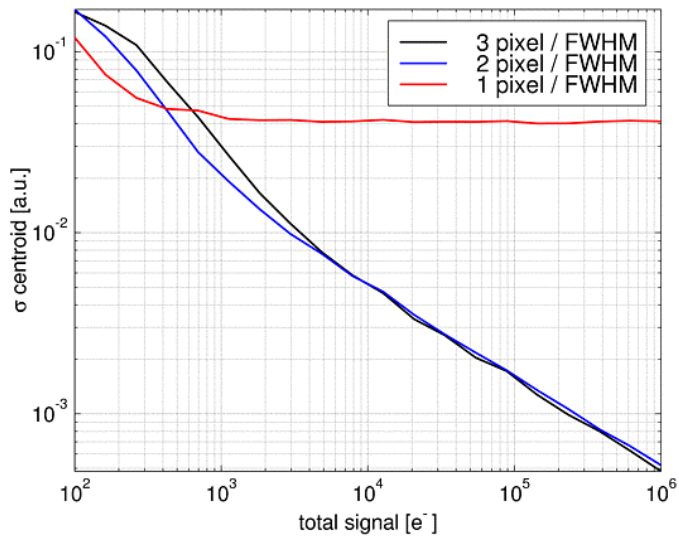

**Figure 7**. Impact of the CCD pixel-scale on the centroiding error.

**Centroiding vs. read-noise**

Figure 8 plots the simulated rms error for the spot centroids against the signal level for different read-noise values. Other simulation parameters have been: 2 pixels / FWHM pixel-scale, WWPA algorithm, 10 x 10 pixels array-size.

As expected, less read-noise provides quite an improvement for very low signals below a few thousand e$^-$. Note that an almost 4 times brighter star is required to achieve an accuracy of 0.1 pixels with a 4e$^-$ read-noise detector compared to a zero read-noise detector.

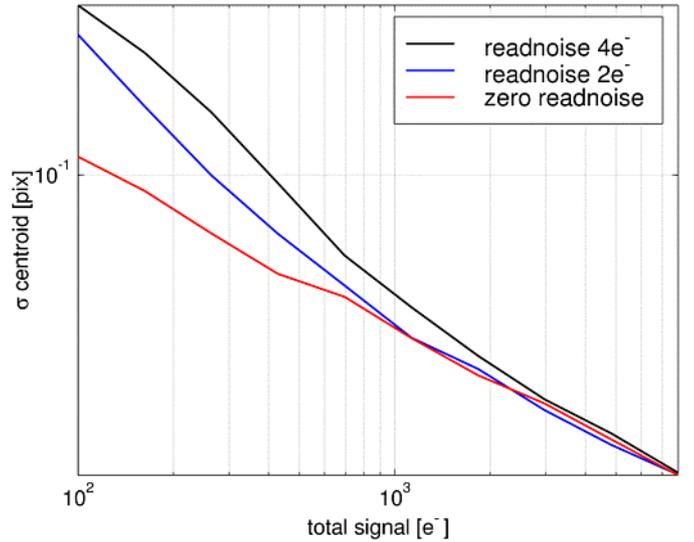

**Figure 8**. Impact of CCD read-noise on the centroiding error.

**4.2. Measurements**

We used a method described by Gendron (1995) to calculate the gradient's measurement noise. The autocorrelation function of the measured gradients shows two features: A smoothly varying component due to the atmospheric turbulence and a spike at the origin due to the uncorrelated measurement noise. The initial value theorem can be used to show that the derivative of the image motion autocorrelation function is zero at the origin. Therefore, a parabola fit can be used to extrapolate to the origin. In practise, taking the difference between the first two values turned out to be a reasonable estimate for the noise variance, see Figure 9.

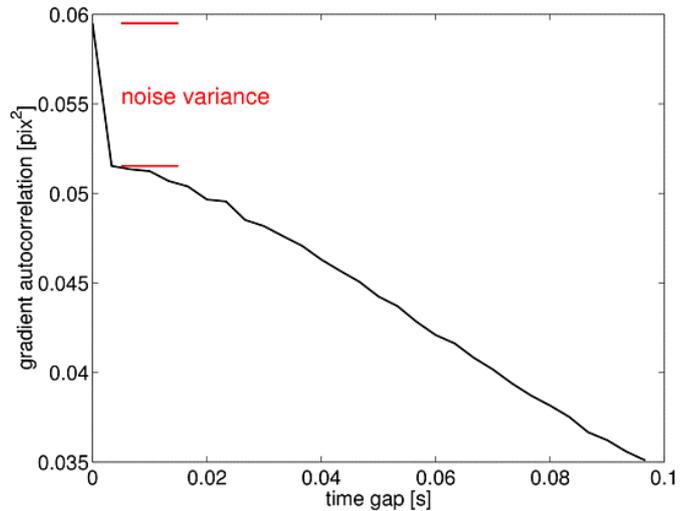

**Figure 9**. Autocorrelation function of measured gradients near the origin.

We implemented the algorithms mentioned in section 4 in the ALFA software, and tested their performance on a star. Sets of 4000 gradients taken with 300 Hz framerate were used throughout the experiment. Besides the different algorithms, 3 different light levels were investigated using neutral density filters. ND0.04 is almost like no filter, ND1.0 has an attenuation factor of 10, ND1.5 has an attenuation factor of 10^1.5 ≈ 31.

The results are shown in Figure 10. The statistical analysis (t-test) indicates that the WWPA algorithm is almost certainly superior to the other algorithms at all light levels. The CWPA and the WPA performances are most sensitive to the signal level, with the WPA performing better at high levels and the CWPA performing better at low levels.

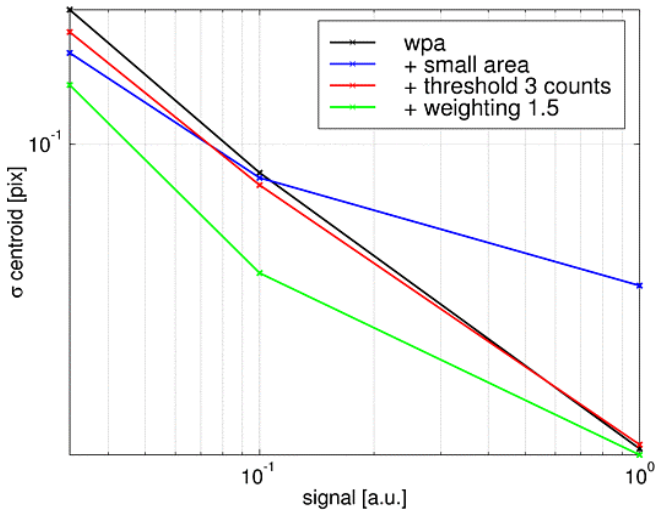

**Figure 10**. Experimental comparison of centroiding algorithms.

## 5. Modal coefficient noise

Knowing the noise covariance of the measured centroids, the noise covariance of the modal coefficients can easily be calculated.

Let the modal coefficients be estimated by a matrix multiplication. So the gradient noise vector $e_g$ propagates accordingly onto the modal coefficients: $e_m = R \cdot e_g$. R is usually called the Reconstruction matrix. The covariance matrix of the modal coefficient noise is then:

$$\langle e_m e_m^t \rangle = R \langle e_g e_g^t \rangle R^t$$

**Wavefront estimators**

Assume the following model for the measured gradients:

$$g = Da + e_g, \quad E \equiv \langle e_g e_g^t \rangle$$

$D$ is the interaction matrix which describes the interaction of modes applied with coefficients $a$ with the sensor measurements.

We investigated two reconstuctors which give an estimate for the modal coefficients based on the noisy measured gradients. These are the commonly used least squares reconstructor: $R_{LS} = (D^t D)^{-1} D^t$, and the weighted least squares reconstructor: $R_{WLS} = (D^t E^{-1} D)^{-1} D^t E^{-1}$ (for additional information see Melsa & Cohn, 1978). $R_{WLS}$ reduces to $R_{LS}$ in case of uniform noise and should perform better, if the noise level varies with the gradients.

Figure 11 shows a measurement for the ALFA D5 array which rules out the assumption of uniform noise. Possible reasons for the observation are unequally illuminated subapertures (see Figure 2) and static aberrations of the microlenses.

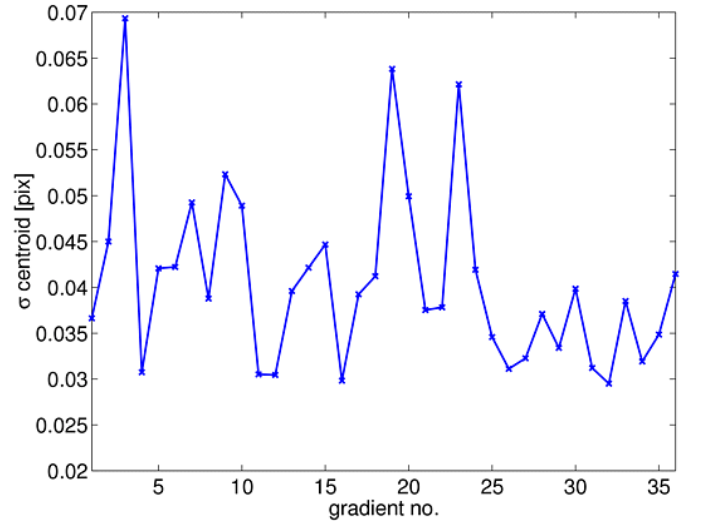

**Figure 11**. Measured noise for the 36 gradients for the D5 array. (18 subapertures can measure 36 x and y gradients).

An experimental comparison of the two reconstructors was made, but under unfavourable seeing conditions (1.3" in K and quite variable). The noise covariance was measured on the star and the reconstruction matrices as introduced above have been used during closed loop operation. The Strehl-ratio was used as a performance indicator because it is a function of the residual wavefront error.

Figure 12 shows that the WLS reconstructor performed better than the LS reconstructor, but the significance level (74%) is not high enough to conclude this with certainty. This experiment will be repeated with more replications.

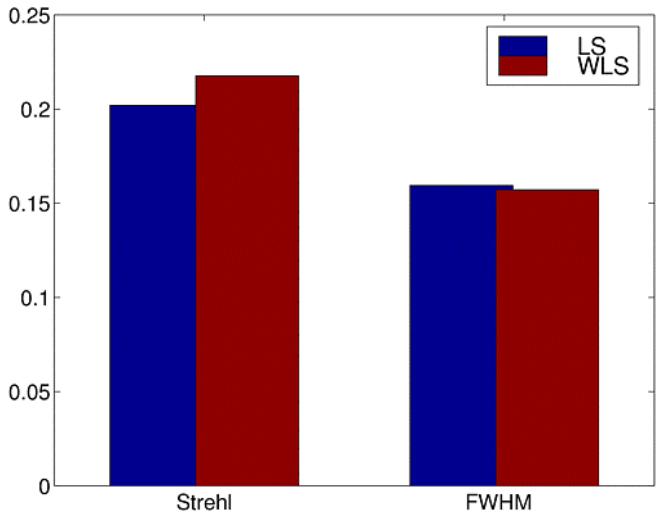

**Figure 12**. Comparison of the Least Squares and the Weighted Least Squares (Gauss-Markov) reconstructors.

**Acknowledgements**. We like to thank Luzma Montoya and Jesús Aceituno from the Calar Alto Observatory for maintaining the ALFA system.